\documentstyle[twoside,fleqn,espcrc2,epsf]{article}

\newcommand{\PR}{Phys.\ Rev.~D }
\newcommand{\PRL}{Phys. Rev. Lett.}

\newcommand{\bea}{\begin{eqnarray}}
\newcommand{\ena}{\end{eqnarray}}

\title{News on Disconnected Diagrams\thanks{Talk presented by J. Viehoff.}}
\author{J.~Viehoff$^{\rm b}$ for the SESAM$^{\rm a,b}$ Collaboration
\\[8pt]
{\small  {\rm $^a$}HLRZ c/o Forschungszentrum J\"ulich, D-52425 J\"ulich,
  and DESY, D-22603 Hamburg, Germany,\\
  {\rm $^b$}Physics Department, University of Wuppertal, D-42097
  Wuppertal, Germany.}}       

\begin{document}

\begin{abstract}
We present evidence for disconnected contributions to the $\sigma_{\pi N}$ term and 
the flavor singlet axial coupling $g_A^0$ of the proton on full QCD configurations, 
which are obtained by means of improved stochastic estimator techniques.

Furthermore we discuss results from the fermionic determination of the
topological charge of the QCD vacuum (in the spirit of the
Athiya-Singer theorem) again achieved with stochastic estimator
methods.  It turns out that this approach provides an 
monitor for the tunneling efficiency of the HMC on QCD with dynamical
Wilson fermions which is independent on the pure gluonic method that implies cooling.
\end{abstract}

\maketitle


\section{Introduction}
Flavor singlet matrix elements, like the $\sigma_{\pi N}$ term
or the axial coupling of the proton $g_A^0$,
have been widely discussed in the past. Recently first lattice calculations 
worked out the connected as well as the disconnected diagrams contributing to
$\sigma_{\pi N}$ and $g_A^0$ \cite{japan,liu}. It turned out that the 
disconnected diagrams define a severe computational problem whereas the methods
for the connected amplitudes work reasonable good. Two different approaches 
to tackle the pure vacuum fluctuations of disconnected diagrams were suggested
in the literature, namely the wall-source method \cite{wall} and stochastic estimator
techniques (SET) \cite{set}. In the following we focus on SET with Z2 noise, coming 
back to the wall-source later on.    


\section{Stochastic Estimator Techniques}
\subsection{Standard Z2 Noise}
Stochastic matrix inversions are based on a set of appropriate source 
vectors $\eta$ where all components carry a Z2 noise and fulfill
\begin{equation} 
\langle \eta_i \rangle = 0, \quad\quad  \langle\eta_i\eta_j\rangle =\delta_{ij}
\end{equation}
in the stochastic average. 
For infinite number of estimates $M^{-1}$ can then be obtained from
\begin{equation}
M^{-1}_{ij}=\langle\eta_j x_i\rangle=\sum_kM^{-1}_{ik}\langle\eta_j\eta_k\rangle
\end{equation}
where $x$ solves $Mx=\eta$ for different source vectors $\eta$. The number of 
estimates equals the number of matrix inversions and the accuracy is bounded according
to the computer facilities available.


\subsection{Improving SET}
The stochastic errors for off-diagonal traces, i.e. $Tr(\gamma_5M^{-1})$, 
$Tr(\gamma_1\gamma_5M^{-1})$, increase drastically  and improved SET methods
become inevitable. Hence, we worked out the effective noise reduction of exact 
inversion in the spin subspace of Wilson's fermion matrix $M$ \cite{latt97}.
However, this method 
reduces the number of estimates for a constant number of inversions by a factor 4,
and could influence the accuracy of $Tr(M^{-1})$ and related observable.
To cope with this problem diagonal-improved SET was invented.
 
The error of $M^{-1}$ for finite numbers of estimates,
\begin{equation}
\Delta(M^{-1}_{ij}) \propto \sum_{k\ne j}M^{-1}_{ik}\langle\eta_j\eta_k\rangle,
\end{equation}
collects a large contribution when $k=i$ due to $M^{-1}_{ii}>>M^{-1}_{ij}$
for the Wilson fermion matrix. The error $\Delta$ can be suppressed by subtracting 
the current estimate for $M^{-1}_{ii}$ after each inversion:
\bea
E[M^{-1}_{ij}]&=& M^{-1}_{ij}+\sum_{k\ne j}M^{-1}_{ik}\langle\eta_j\eta_k\rangle\nonumber\\
               && - E[M^{-1}_{ii}]\langle\eta_j\eta_i\rangle
\ena
where $i\ne j$. This {\it diagonal-improved} SET reduces the standard 
error for $Tr(\gamma_5M^{-1}), Tr(\gamma_1\gamma_5M^{-1})$ and $Tr(\gamma_2\gamma_5M^{-1})$ 
about $40\%$ without additional matrix inversions and does not reflect on $Tr(M^{-1})$.


\section{Flavor Singlet Matrix Elements}
\begin{figure}[htb]
\begin{center}
\noindent\parbox{7.6cm}{
\parbox{7cm}{\epsfxsize=7cm\epsfysize=5cm\epsfbox{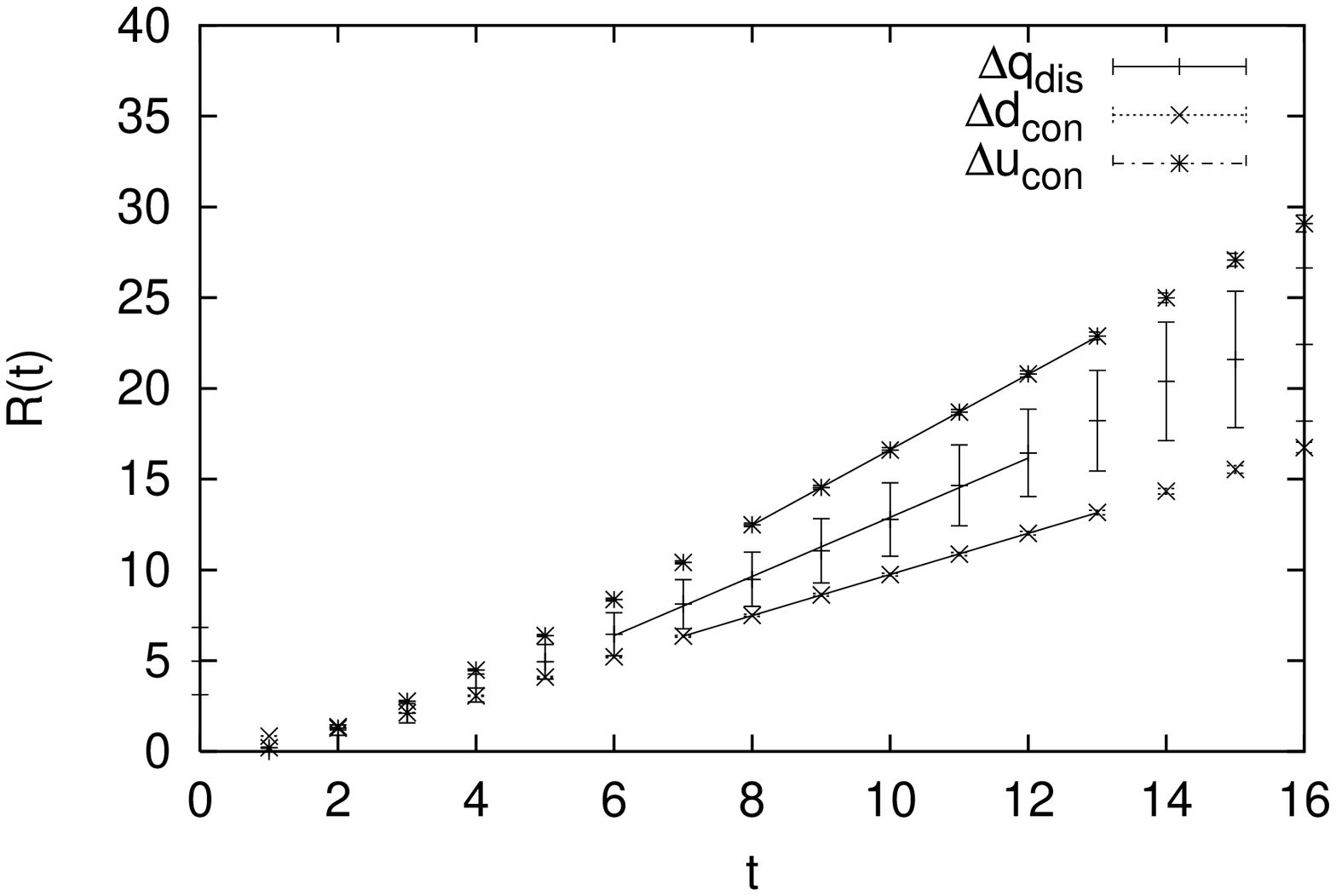}}}
\vskip -0.7cm
\noindent\parbox{7.6cm}{
\parbox{7cm}{\epsfxsize=7cm\epsfysize=5cm\epsfbox{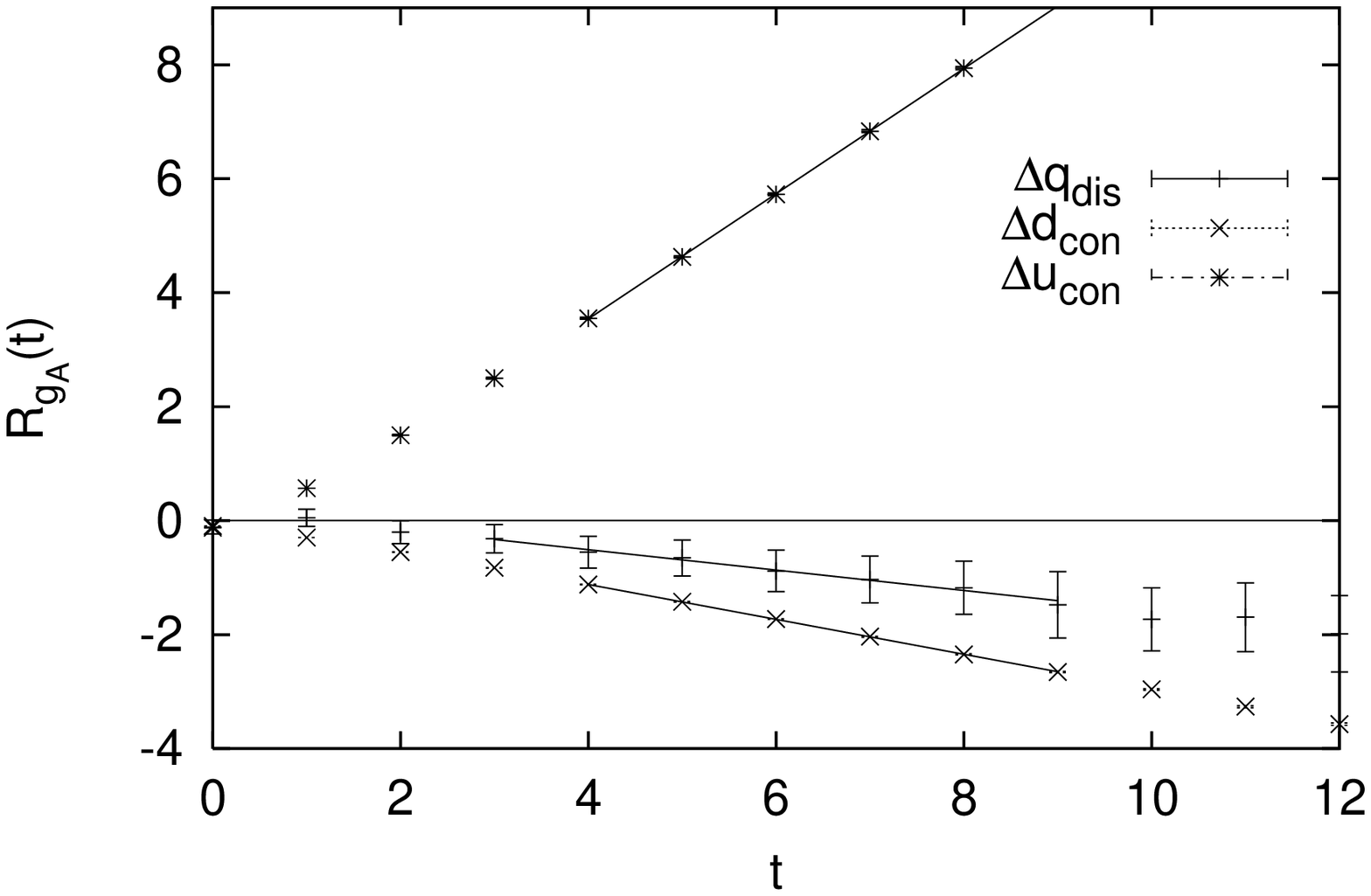}}}
\end{center}
\vskip -0.7cm
\caption[a]{\label{fig_ratio}
\it $R(t)$ and $R_{g_A}(t)$ for the scalar and axial insertion on 200 configurations,
$\kappa=0.156$. 
The linear slopes are proportional to $\Delta u_{con}$, $\Delta d_{con}$ and
$\Delta q_{dis}$.} 
\end{figure}

We look at matrix elements of the proton 
$\langle p|\bar{q}\Gamma q|p\rangle = \Delta q$
at zero momentum when $\bar{q}\Gamma q$ is a flavor singlet bi-quark operator, i.e. 
$\Gamma = 1$ for the $\sigma_{\pi N}$ term and $\Gamma = \gamma_5\gamma_{\mu}$
for the axial coupling $g_A^0$. The ratio of the three-point function and the
proton propagator 
\bea
R(t)&=&\frac{\langle p(t) (\sum_n\bar{q}\Gamma q) \bar{p}(0)\rangle}{\langle p(t)\bar{p}(
0)\rangle}\nonumber\\
\nonumber\\
&&\stackrel{t\gg 1}{\longrightarrow} const +Z_{\Gamma}^{-1}\langle p|\bar{q}\Gamma q | p\rangle \,t\; ,
\label{ratio}
\ena
is for large $t$ proportional to $\Delta q$ \cite{maiani_ratio}. Connected diagrams are 
calculated with the conventional source method \cite{source} whereas for the 
disconnected amplitudes we applied improved SET to estimate $Tr(\Gamma M^{-1})$ 
which is required in eq. \ref{ratio}. The SESAM gauge-field configurations are generated 
with 2 flavors of dynamical fermions on $16^3\times 32$ lattices with the 
standard Wilson action ($\beta =5.6$, $\kappa =0.1560-0.1575$). 
Figure \ref{fig_ratio} shows $R(t)$ on 200 
configurations at $\kappa =0.156$ for  $\sigma_{\pi N}$ and $g_A^0$,
respectively. For scalar insertions the disconnected amplitudes are compatible
to the connected ones. $\Delta u_{con}, \Delta d_{con}$ and 
$\Delta q_{dis}=\Delta u,d_{dis}$ are positive. Axial connected insertions have
opposite sign and $\Delta q_{dis}$ is rather small and negative. Even with
improved SET the errors for $\Delta q_{dis}$ are still around 30\% . 
The outstanding calculation of $\Delta s_{dis}$ for the strange quark will
complete the analysis for $\sigma_{\pi N}$ and $g_A^0$ and is in 
advanced progress. Finally we should point out that the wall-source method is too
noisy for the axial disconnected insertion and improved SET allows the
determination of $g_A^0$ on our set of full QCD configurations.


\section{Topology and SET}
\begin{figure}[htb]
\begin{center}
\noindent\parbox{7.6cm}{
\parbox{7cm}{\epsfxsize=7cm\epsfysize=5cm\epsfbox{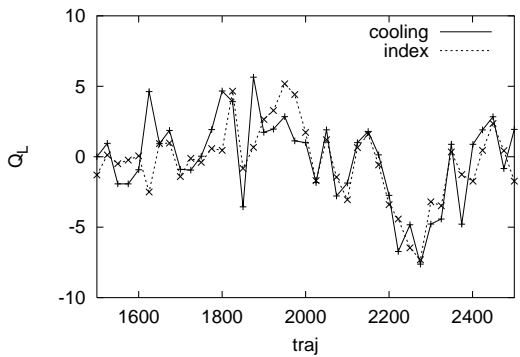}}}
\end{center}
\vskip -0.7cm
\caption[a]{\label{topol}
\it The field-theoretical definition of the topological charge $Q_L$ after cooling 
and via the index theorem on a set of HMC trajectories for $\kappa=0.1575$.}
\end{figure}

Topological properties of the QCD vacuum are essential for the 
understanding of flavor singlet physics, e.g. the unexpectedly large $\eta'$ mass 
or the proton spin. Here the topological charge $Q$ plays a crucial role since the 
Witten-Veneziano formula relates the topological susceptibility, 
$\chi = \langle Q\rangle / V$, directly to
the $\eta'$ mass. Instead of using the more usual field-theoretic \cite{field} 
or geometric definitions \cite{geo} of $Q$, we look at the Athiya-Singer 
index theorem in the continuum 
\begin{equation}
Q = n_+ - n_- = m_q Tr(\gamma_5 G(x,x'))
\end{equation}
where $n_{+-}$ are the numbers of left and right handed chiral eigen-modes, $m_q$
the quark mass of the propagator $G(x,x')$. On the lattice this turns into 
\begin{equation}
\label{index}
Q_L=m_q \kappa_P Tr(\gamma_5M^{-1}). 
\end{equation}
Here $M^{-1}$ is the inverse fermion 
matrix and $\kappa_p$ a renormalization constant \cite{vink}. Again we used improved 
SET to estimate the right hand side of eq. \ref{index}. In figure \ref{topol} the 
results for $Q_L$ on a set of HMC trajectories from SESAM lattices 
($\kappa =0.1575$) are displayed. The topological charge via the index theorem
is compared to $Q_L$ when we use cooling and the field-theoretical definition \cite{field}.
The best overlap of both data is achieved by setting 
$m_q \kappa_P = 2 m_q$ in eq. \ref{index} and a further determination of
$\kappa_P$ is imperative. The strong correlation of the data indicates that the 
index theorem remains valid on the lattice. Finally two completely different
methods display the tunneling efficiency of the HMC algorithm through various
topological sectors and justify the ongoing calculation of $\chi$, the $\eta'$ mass 
and the proton spin. 

\section{Conclusion and Outlook}
We presented an improved stochastic estimator technique and found 
evidence that disconnected diagrams contribute to the $\sigma_{\pi N}$ term and to 
the flavor singlet axial coupling of the proton, $g_A^0$. It turns out that the
disconnected contribution $\Delta u,d_{dis}$ to $\sigma_{\pi N}$ is comparable to 
the connected amplitude, whereas $\Delta q_{dis}$ for $g_A^0$ is rather small
but non-vanishing. The final analysis for both flavor singlet operators, 
$\sigma_{\pi N}$ and $g_A^0$, will be published soon.

Furthermore we achieved values for the topological charge $Q_L$ with stochastic 
matrix inversion. These data are in good agreement with the results for $Q_L$
when the field-theoretical definition and cooling is applied.  
Both methods demonstrate the tunneling efficiency  of the HMC on SESAM 
lattices throughout different topological sectors.


\end{document}